\documentclass[aps,prl,reprint,twocolumn,amsmath,amssymb,superscriptaddress,showpacs]{revtex4-1}

\usepackage{graphicx}
\usepackage{bm}
\usepackage{color}
\usepackage{hyperref}
\begin{document}
	
	\title{Attractive Coulomb interactions in a triple quantum dot}%
	
	\author{Changki Hong}
	\affiliation{Department of Physics, Pusan National University, Busan 46241, Republic of Korea}
	\author{Gwangsu Yoo}
	\affiliation{Department of Physics, Korea Advanced Institute of Science and Technology, Daejeon 34141, Republic of Korea}
	\author{Jinhong Park}
	\affiliation{Department of Condensed Matter Physics, Weizmann Institute of Science, Rehovot 76100, Israel}
	\author{Min-Kyun Cho}
	\affiliation{Department of Physics and Astronomy, and Institute of Applied Physics, Seoul National University, Seoul 08826, Republic of Korea}
	\author{Yunchul Chung}
	\email{ycchung@pusan.ac.kr}
	\affiliation{Department of Physics, Pusan National University, Busan 46241, Republic of Korea}
	\affiliation{Department of Condensed Matter Physics, Weizmann Institute of Science, Rehovot 76100, Israel}
	\author{H.-S. Sim}
	\email{hs\_sim@kaist.ac.kr}
	\affiliation{Department of Physics, Korea Advanced Institute of Science and Technology, Daejeon 34141, Republic of Korea}
	\author{Dohun Kim}
	\email{dohunkim@snu.ac.kr}
	\affiliation{Department of Physics and Astronomy, and Institute of Applied Physics, Seoul National University, Seoul 08826, Republic of Korea}
	\author{Hyungkook Choi}
	\affiliation{Department of Physics, Research Institute of Physics and Chemistry, Chonbuk National University, Jeonju 54896, Republic of Korea}
	\author{Vladimir Umansky}
	\affiliation{Department of Condensed Matter Physics, Weizmann Institute of Science, Rehovot 76100, Israel}
	\author{Diana Mahalu}
	\affiliation{Department of Condensed Matter Physics, Weizmann Institute of Science, Rehovot 76100, Israel}
	\begin{abstract}
		Electron pairing due to a repulsive Coulomb interaction in a triple quantum dot (TQD) is experimentally studied. It is found that electron pairing in two dots of a TQD is mediated by the third dot, when the third dot strongly couples with the other two via Coulomb repulsion so that the TQD is in the twofold degenerate ground states of $(1,0,0)$ and $(0,1,1)$ charge configurations. Using the transport spectroscopy that monitors electron transport through each individual dot of a TQD, we analyze how to achieve the degeneracy in experiments, how the degeneracy is related to electron pairing, and the resulting nontrivial behavior of electron transport. Our findings may be used to design a system with nontrivial electron correlations and functionalities.
	\end{abstract}
	\maketitle
	
	Recently, it was experimentally demonstrated \cite{Hamo2016}, using an electrostatically coupled quadruple quantum dot formed in carbon nanotubes, that an effectively attractive interaction between electrons can be induced purely by Coulomb repulsion. An attractive interaction occurs between two electrons in two dots of a quadruple quantum dot, with the help of the other two dots \cite{Raikh1996}. This mechanism, called an "excitonic" attraction, was originally proposed by Little \cite{Little1964} as a possible pairing mechanism to engineer high-$T_C$ superconductivity in an organic superconductor. 
	
	In this work, we experimentally show that an attractive interaction can be also realized in a simpler system of a triple quantum dot (TQD). When a TQD has twofold degenerate ground states of $(n_1, n_2, n_3) = (1,0,0)$ and $(0,1,1)$ charge configurations, $n_i$ being the electron occupation number of QD$i$ in the TQD, QD2 and QD3 prefer the occupancy $(n_2, n_3) = (1,1)$ of an electron pair or zero occupancy $(0,0)$, while single-electron occupation, $(1,0)$ or $(0,1)$, is blocked by its repulsive interaction with QD3. Although this two-fold degeneracy in a TQD was observed previously \cite{Gaudreau2006,Granger2010,Rogge2008,Rogge2009,Schroer2007}, its properties, such as the connection to electron pairing, how to achieve it, and its signatures in electron transport, have not been studied. It is partly because it is difficult to control the coupling between the QDs separately and it is too demanding to measure the full three-dimensional stability diagram \cite{Schroer2007} of a TQD than the two-dimensional stability diagram of a double quantum dot. Here, we study electron transport through a TQD. In our setup (see Fig. 1), it is possible to separately tune electrostatic coupling between the QDs of the TQD and to monitor electron transport through each QD, in contrast with previous studies \cite{Gaudreau2006,Rogge2008}. This transport spectroscopy allows us to analyze the three-dimensional stability diagram of the TQD, the condition for the degeneracy, and the connection of the degeneracy with electron pairing. Based on the concept of effective charges, we find that electron pairing in two QDs of the TQD is mediated by the third QD, when the third QD strongly couples with the other two QDs via Coulomb repulsion. This results in characteristic transport behavior. Our findings will be useful for designing multiple quantum dot systems with nontrivial electron correlations and functionalities \cite{Hamo2016,Yoo2014}.
	\begin{figure}[h]
		\begin{center}
			\includegraphics[width=1\linewidth]{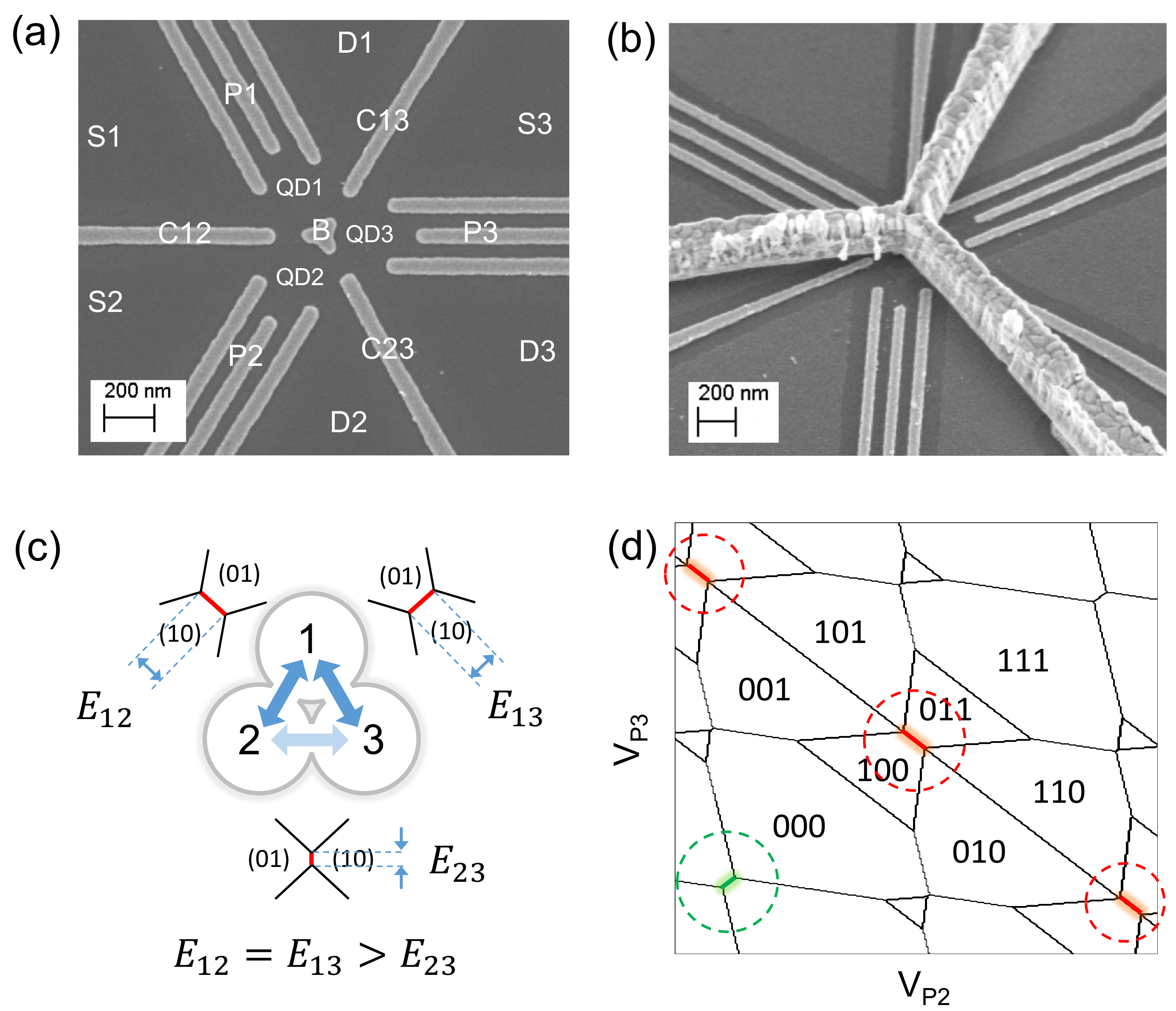}
			\caption{(color online) (a) A scanning electron microscope (SEM) image of a similar device used in the experiment. The sample was fabricated on a GaAs/AlGaAs heterostructure wafer having a 78.5 nm deep two-dimensional electron gas (2DEG) layer with carrier density $n=2.2\times10^{11}$ cm$^{-2}$ and mobility $\mu=4.7\times10^{6}$ cm/V s. (b) The final device image with a metallic bridge floating over the QD structure to connect the island gate B. (c) To achieve the degeneracy for electron pairing, $E_{12}$ and $E_{13}$ are kept stronger than $E_{23}$, where $E_{ij}$ is the electrostatic coupling energy between QD$i$ and QD$j$. (d) Stability diagram as a function of $V_{\textrm{P}2}$ and $V_{\textrm{P}3}$, where $V_{\textrm{P}i}$ is the voltage applied to the plunger gate of QD$i$. $V_{\textrm{P}1}$ is fixed at a value with which the degeneracy is obtained. It is computed for a TQD with the symmetry between QD2 and QD3, using Eq.~(1). Electron occupation numbers are labeled as $(0,0,0)$, for clarity, by subtracting constants from the actual numbers. The actual numbers are roughly larger than 100. The degeneracy for electron pairing occurs between $(1,0,0)$ and $(0,1,1)$ (see a red circle in the center), between $(1,-1,1)$ and $(0,0,2)$ (the upper left-hand corner), and between $(1,1,-1)$ and $(0,2,0)$ (the lower right-hand corner).
			}
			\label{fig:short}
		\end{center}
	\end{figure}

	Figure 1(a) is an image of our device. Three QDs are separated by an island gate B and coupling gates C12, C13, and C23, enabling one to tune coupling between the QDs separately. As each QD has its own source and drain, electron conductance through each QD is measured in our transport spectroscopy. The island gate is connected by a metallic bridge floating over the device. To make capacitive coupling between the bridge and the QDs symmetric (this allows us to tune the QDs easily), the bridge was designed symmetrically as in Fig. 1(b). 
	
	The TQD can be described by a constant interaction model. Its electrostatic energy $U(n_1,n_2,n_3)$ is written as \cite{Schroer2007}
	\begin{equation} \label{eq : 1}
	U(n_1,n_2,n_3) = \frac{1}{2e^2}\sum_{i,j=1,2,3}E_{ij}Q_iQ_j.
	\end{equation}
	$E_{ii}$ is the onsite Coulomb energy of QD$i$ and $E_{ij(\ne{i})}$ is the Coulomb interaction between QD$i$ and QD$j$. $E_{ij}$'s are written in terms of capacitances. $Q_i=-en_i+\sum_{\sigma=1,2,3}D_{i\sigma}V_{\textrm{P}\sigma}$ is the effective charge of QD$i$, where $D_{i\sigma}$ is the mutual capacitance between QD$i$ and plunger gate $\sigma$ of voltage $V_{\textrm{P}\sigma}$, and $e$ is the electron charge. Using Eq. (1), we find that the two-fold degenerate ground states of $(n_1,n_2+1,n_3+1)$ and $(n_1+1,n_2,n_3)$ can appear when $E_{12},E_{13}>E_{23}$ [see Fig. 1(c)], or equivalently when the QD2-QD3 capacitance is smaller than the QD1-QD2 and QD1-QD3 capacitances; when $E_{12}=E_{13}=E_{23}$, a sixfold degeneracy for charge frustration occurs instead \cite{Seo2013}. Under the condition and the model in Ref. \cite{Schroer2007}, we draw the stability diagram in Fig.~1(d). It shows the twofold degeneracy for electron pairing (EP degeneracy) along the red line, where QD2 and QD3 prefer the occupancy $(n_2,n_3)=(1,1)$ of an electron pair or zero occupancy $(0,0)$ rather than single occupancy $(1,0)$ or $(0,1)$. Around the EP degeneracy line (see the red circle), the diagram is dramatically different from that of a double quantum dot (DQD) \cite{Wiel2002}, while it is similar around the DQD degeneracy line (the green circle) of $(0,-1,0)$ and $(0,0,-1)$. 
	\begin{figure}[h]
		\begin{center}
			\includegraphics[width=1\linewidth]{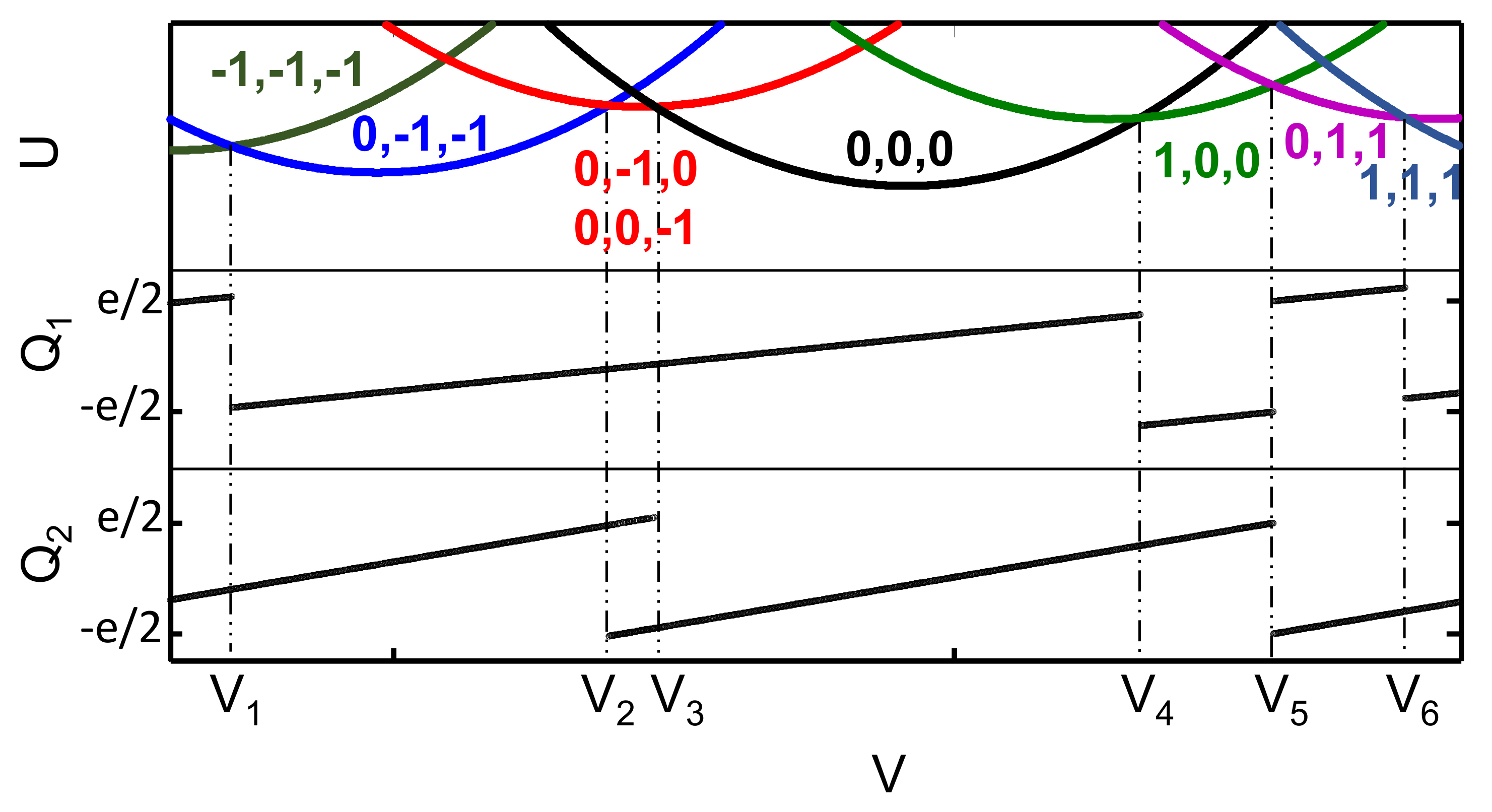}
			\caption{(color online) In the upper panel, the occupation-number state $(n_1,n_2,n_3)$ having the lowest electrostatic energy $U$ is presented along the line of $V \equiv$$a{V}_{\textrm{P}1}=V_{\textrm{P}2}=V_{\textrm{P}3}$. $a\approx2.04$ is chosen such that the line passes through the center of a domain of $(V_{\textrm{P}1},V_{\textrm{P}2},V_{\textrm{P}3})$ in which the states $(1,0,0)$ and $(0,1,1)$ are the degenerate ground states of $U$. For each $(n_1,n_2,n_3)$, $U(n_1,n_2,n_3)$ is plotted (parabolas) as a function of $V$. Dashed lines indicate the values of $V$ at which a transition between different ground states occurs.  The effective charges $Q_1$ (middle panel) and $Q_2$ (lower) are also plotted as a function of $V$. This plot is drawn with the same parameters as Fig. 1(d).}
			\label{fig:short}
		\end{center}
	\end{figure}
	
	To understand the EP degeneracy, we analyze the effective charges for a left-right symmetric TQD of $E_{22}=E_{33}$ and $E_{12}=E_{13}>E_{23}$. Figure 2 shows the ground-state energy of $U(n_1,n_2,n_3)$ and the effective charges $Q_i$ as a function of plunger gate voltage $V$. At $V=V_1$, the QD1 degeneracy of $(-1,-1,-1)$ and $(0,-1,-1)$ occurs. At this point, $Q_1$ is $e/2$ or $-e/2$, causing sequential electron tunneling through QD1. QD1 degeneracy also happens at $V=V_4$ and $V_6$.
	
	In $V_2<V<V_3$, the DQD degeneracy of $(0,-1,0)$ and $(0,0,-1)$ occurs. Here, $Q_1 \simeq 0$ so that QD1 is electrostatically inert, and $(Q_2,Q_3)$ has the two values of $Q_2=-Q_3\simeq \pm{e}/2$. To reduce the repulsive interaction between QD2 and QD3, $(0,-1,0)$ and $(0,0,-1)$ become the ground states. This DQD degeneracy corresponds to the green line of the stability diagram in Fig. 1(d), and to the $(1,0)$ and $(0,1)$ degeneracy of a DQD \cite{Wiel2002}. The length of the line is proportional to $E_{23}$. Along the degeneracy line, cotunneling \cite{Bruus2004} and an orbital Kondo effect can occur \cite{Amasha2013}.
	
	At $V=V_5$, the EP degeneracy of $(1,0,0)$ and $(0,1,1)$ occurs, and $Q_i$'s have the two values of $Q_1=-Q_2=-Q_3=\pm{e}/2$. The energy cost for the electron pairing, Coulomb repulsion between QD2 and QD3, is compensated by electron-hole attractive interactions between QD1 and QD2 and between QD1 and QD3 when $E_{12},E_{13}>E_{23}$. It happens as follows. When the effective charge $Q_1$ of QD1 is $-e/2$, an electron is pushed out of each of QD2 and QD3 to make its effective charge $+e/2$. When $Q_1$ is $+e/2$, an electron is pulled into each of QD2 and QD3 to make its effective charge $-e/2$. These processes effectively make the pairing of one electron in QD2 and another in QD3. This EP degeneracy point of $V=V_5$ corresponds to the red line of the stability diagram in Fig. 1(d). The length of the line is proportional to $E_{12}-E_{23}$. Along the line, a transition between the two ground states occurs via high-order cotunneling in which all the three QDs are involved, and an anisotropic charge Kondo effect can occur, when the TQD couples to reservoirs via electron tunneling \cite{Yoo2014}.
	
	The EP degeneracy has a number of interesting features. First, the EP degeneracy line of $(1,0,0)$ and $(0,1,1)$ is almost orthogonal to the DQD degeneracy line of $(0,-1,0)$ and $(0,0,-1)$ as in the stability diagram in Fig. 1(d). Second, around the EP degeneracy line, the effective charge $Q_1$ of QD1 is accumulated with a jump from $-e/2$ to $+e/2$ as $V$ increases (see Fig. 2). This is opposite to the QD1 degeneracy where $Q_1$ is relaxed with a jump from $+e/2$ to $-e/2$ as $V$ increases. To compensate this unnatural charge accumulation, an EP degeneracy line is accompanied by two QD1 degeneracy lines; see the degeneracy of $(0,0,0)$ and $(1,0,0)$ and that of $(0,1,1)$ and $(1,1,1)$ in Fig. 1(d), and the degeneracy at $V=V_4$ and $V_6$ in Fig. 2. Third, along the EP degeneracy line in Fig. 1(d), the ground states are $(1,0,0)$ and $(0,1,1)$. The next lowest excitations are $(1,0,1)$, $(1,1,0)$, $(0,1,0)$, and $(0,0,1)$, where the effective charge $Q_1$ of QD1 repels one of $Q_2$ and $Q_3$. The highest-energy charge configurations are $(0,0,0)$ and $(1,1,1)$ where $Q_1$ repels both of $Q_2$ and $Q_3$.
	\begin{figure}[h]
		\begin{center}
			\includegraphics[width=1\linewidth]{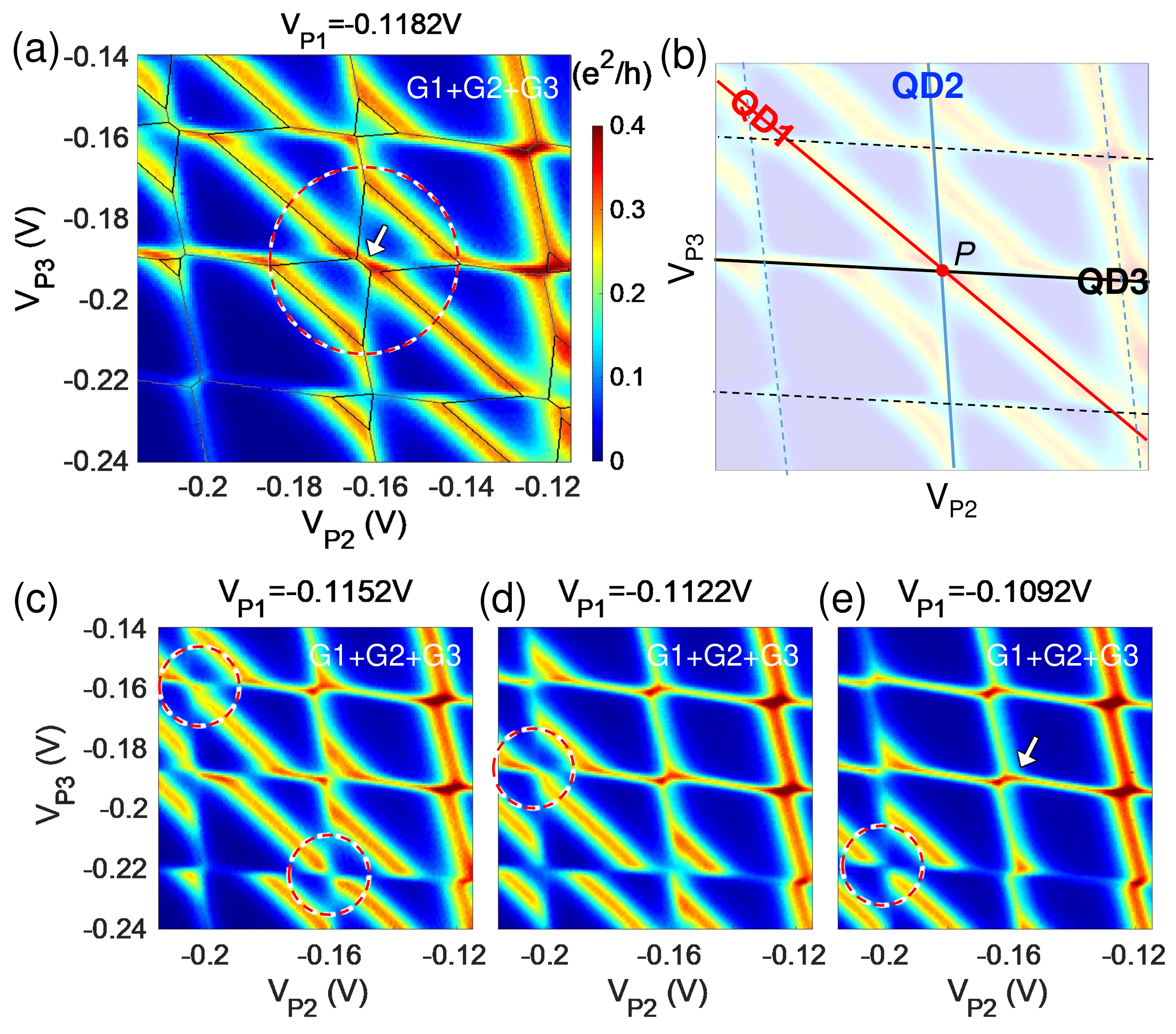}
			\caption{(color online) Stability diagram measured with varying $V_{\textrm{P}2}$ and $V_{\textrm{P}3}$, at fixed $V_{\textrm{P}1}$ at (a) $-0.118$, (c) $-0.115$, (d) $-0.112$, and (e) $-0.109$ V. The black lines in (a) indicate the stability diagram calculated with the capacitances in Ref. \cite{Fit}. The conductance measured only through QD2 and QD3 (G2+G3) is presented in the Supplemental Material \cite{Supplemental}. (b) Dashed lines represent the charge degeneracy lines of QD1 (red), QD2 (blue), and QD3 (black). For all the figures, color scales are the same as the one in (a).}
			\label{fig:short}
		\end{center}
	\end{figure}

	We now discuss experiments, which confirm the above analysis. Measurements were done in a dilution refrigerator of base temperature 10 mK (electron temperature $<$ 100 mK). Electron conductance through the TQD was measured by applying 750Hz, 10$\mu$V$_{\textrm{rms}}$ AC modulation voltage to the sources and measuring electric current at the drains by using a home-made wideband current amplifier \cite{Andrey2012} combined with a lock-in amplifier. Zero-bias electron conductance G$i$ through an individual dot QD$i$ from its source to its drain is measured in our setup. This allows us to get more information than previous experiments \cite{Hamo2016,Gaudreau2006,Rogge2008}, hence, to unveil the features of the EP degeneracy.
	
	The TQD is formed as follows. First, the coupling between two dots (QD1-QD2, QD1-QD3, QD2-QD3) are adjusted to satisfy the electron pairing condition $E_{12} (\sim 180\mu{eV}) \approx{E}_{13} (\sim 160\mu{eV}) >E_{23}(\sim 70 \mu{eV})$, by measuring the stability diagram of the two dots and tuning the coupling gate C12, C13, C23 [see Fig. 1(c)]. Then, the stability diagram of the TQD is measured by adding conductance G1, G2, and G3 (that we call G1+G2+G3). In Figs. 3(a) and (c-e), the stability diagram is plotted at various QD1 plunger gate voltages V$_{\textrm{P}1}$ with varying V$_{\textrm{P}2}$ and V$_{\textrm{P}3}$ [see the red circles in Figs. 3(c)-3(e) and the video in the Supplemental Material \cite{Supplemental}]. 
	
	Figure 3(a) shows that the measured stability diagram matches well with a calculation result (black line), especially around the central region (inside the red dashed circle) where the EP degeneracy of (1,0,0) and (0,1,1) is expected. The horizontal, vertical and diagonal lines in Fig. 3(b) roughly corresponds to the charge degeneracy lines of QD3, QD2, and QD1, respectively, in the case of no (or little) interaction between QDs. These three lines are crossing at the center $P$ [or the white arrow in Fig.~3(a)] of the EP degeneracy line of (1,0,0) and (0,1,1). This implies that the effective charge of each QD is close to $\pm{e}/2$ along the EP degeneracy line, and that the interaction between QD2 and QD3 is effectively attractive, as discussed in Fig. 2. As the QD1 plunger gate voltage changes, the effective charge of QD1 becomes close to zero so that the interaction between QD2 and QD3 becomes repulsive. This is confirmed in the analysis of the three-dimensional stability diagram on the $(V_{\textrm{P}1},V_{\textrm{P}2},V_{\textrm{P}3})$ space in Figs.~3(c-e). As V$_{\textrm{P}1}$ changes, the diagonal charge degeneracy line of QD1 recedes from point $P$, and the DQD degeneracy line instead appears at the point $P$ in Fig. 3(e).
	
	By changing V$_{\textrm{P}1}$, it was possible to observe EP degeneracy at every possible lattice points of the stability diagram (see red circles in Figs. 3(c-e) and the supplement video) where a vertical (QD2) and a horizontal (QD3) charge degeneracy line cross a diagonal (QD1) charge degeneracy line. This shows that the EP degeneracy happens consistently under the proper coupling between the QDs $(E_{12}\approx{E}_{13}>E_{23})$, regardless of the total number of electrons and electron spins in the QDs.
	\begin{figure}[h]
		\begin{center}
		\includegraphics[width=1\linewidth]{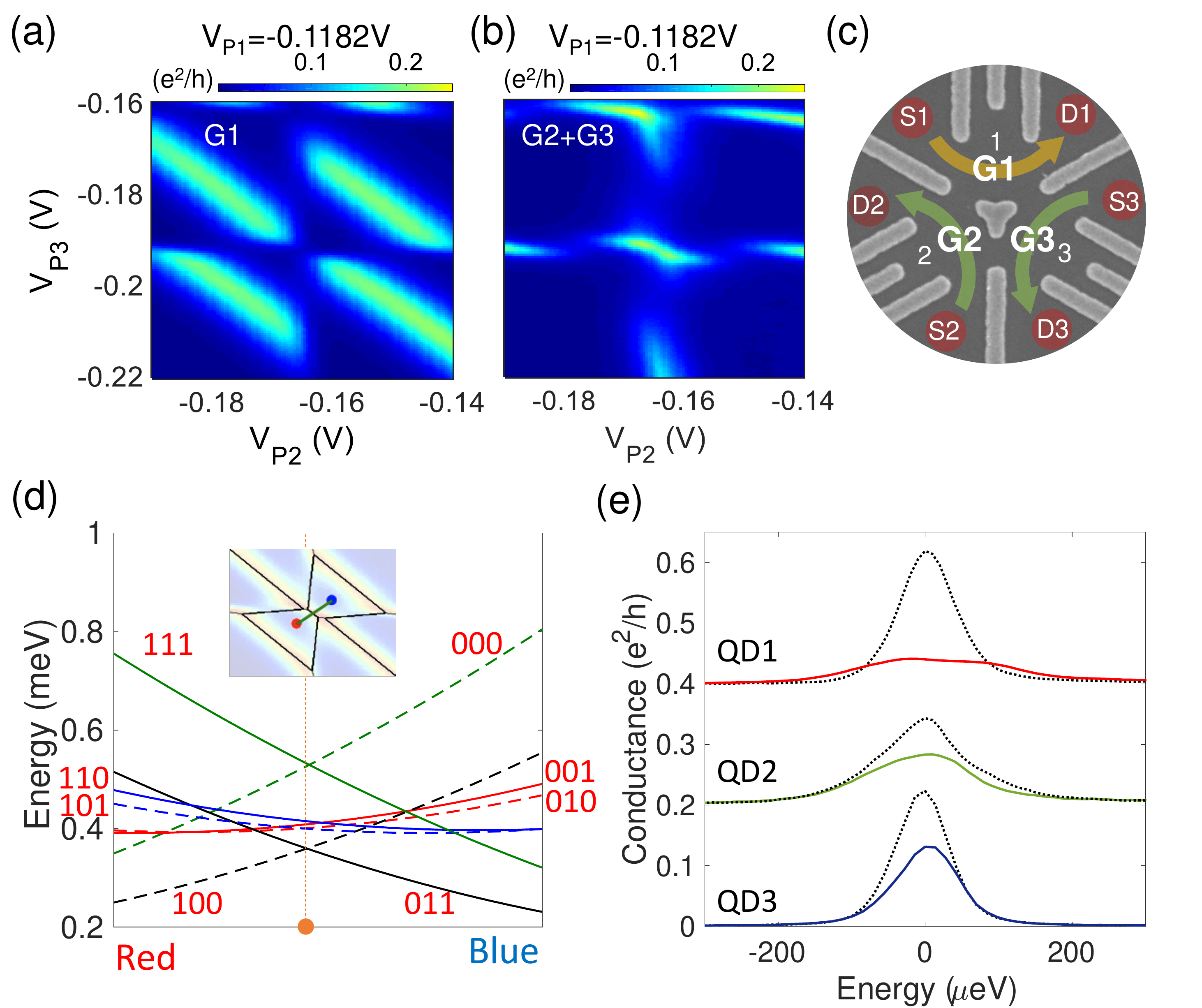}
		\caption{(color online) Electron conductance through (a) QD1 only (G1), and that through (b) QD2 and QD3 (G2 + G3), measured around the central region of Fig. 3(a). (c) Conductance through QD1, QD2, QD3 can be measured as G1, G2, G3, respectively. (d) Energy levels of the TQD around the EP degeneracy line, calculated with the same capacitances as those used for calculating the stability diagram in Fig. 3(a). The x-axis of the plot is scanned from the red point to the blue point in the inset. The orange point indicates the center of the EP degeneracy line of (0,1,1) and (1,0,0). (e) Conductance peaks (solid lines) of G$i$'s measured along the line in the inset of (d). The peak center occurs at the EP degeneracy point. A conductance peak (dotted black lines) measured around a charge degeneracy line of each QD (e.g., around that of (0,0,0) and (1,0,0) for QD1) is shown for comparison. The voltage is converted into energy scale $\alpha{V}_P$ by measuring $\alpha$ \cite{Kouwenhoven2001} for each QD. The peaks are horizontally shifted such that their centers are located at the same position. Also, for clarity, the peaks of QD1 and QD2 are shifted vertically by $0.4e^2/h$ and $0.2e^2/h$, respectively.}
			\label{fig:short}
		\end{center}
	\end{figure}
	
	In Figs.~3(a) and 1(d), the EP degeneracy line of (1,0,0) and (0,1,1) is connected with the QD1 degeneracy lines of (0,0,1) and (1,0,1) and of (0,1,0) and (1,1,0). They are distinguished in our transport spectroscopy shown in Fig.~4(c). In Figs.~4(a) and 4(b), conductance (G1) through QD1 and that (G2+G3) through QD2 and QD3 are separately measured. Along the QD1 degeneracy line, G1 has relatively large values, while G2+G3 is negligibly small. This is consistent with the fact that the effective charge $Q_1$ of QD1 can fluctuate at the QD1 degeneracy, while $Q_2$ and $Q_3$ cannot. 
	On the other hand, along the EP degeneracy line, G2+G3 has relatively large values, while G1 is very small. 
	
	The behavior on the EP line in Figs.~4(a) and 4(b) can be understood based on the excitation energy spectrum in Fig.~4(d) and conductance peak broadening in Fig.~4(e). When the TQD weakly couples to the sources and drains via electron tunneling, conductance is mediated by cotunneling processes along the EP degeneracy line.  In this case, both  G1 and G2+G3 are not as large as the measured result in Fig.~4. We find that the coupling of our TQD to the leads is so strong that the peak broadening of G1, G2, G3 is 100$\mu$eV, 120$\mu$eV and 80$\mu$eV, respectively [see Fig.~4(e)]. The peak broadenings are larger than the excitation energy $\Delta{E}_1\sim45\mu{eV}$ to the first excited states of (1,0,1), (1,1,0), (0,1,0), (0,0,1), but smaller than the excitation energy $\Delta{E}_2\sim170\mu{eV}$ to the second excited states (0,0,0) and (1,1,1). Then the first excited states can contribute to G2+G3 via sequential tunneling processes of (0,1,1)$\to$(0,1,0), (0,1,1)$\to$(0,0,1), (1,0,0)$\to$(1,0,1), and (1,0,0)$\to$(1,1,0). However, the sequential processes contributing to G1, such as (0,1,1)$\to$(1,1,1) and (1,0,0)$\to$(0,0,0), are not allowed and only weak cotunneling processes \cite{Bruus2004} contribute to G1, since $\Delta{E}_2$ is larger than the peak broadening and the electron temperature. This behavior of G2+G3 $\gg$ G1 along the EP line is relevant to the attractive interaction: The energy cost for charge fluctuations through the site (QD1) inducing the attractive interaction is larger than that for charge fluctuations through the sites (QD2 and QD3) where the attractive interaction occurs. In this way, the attractive interaction is stabilized under the condition of $E_{12} \simeq E_{13} > E_{23}$.
	\begin{figure}[h]
		\begin{center}
		\includegraphics[width=1\linewidth]{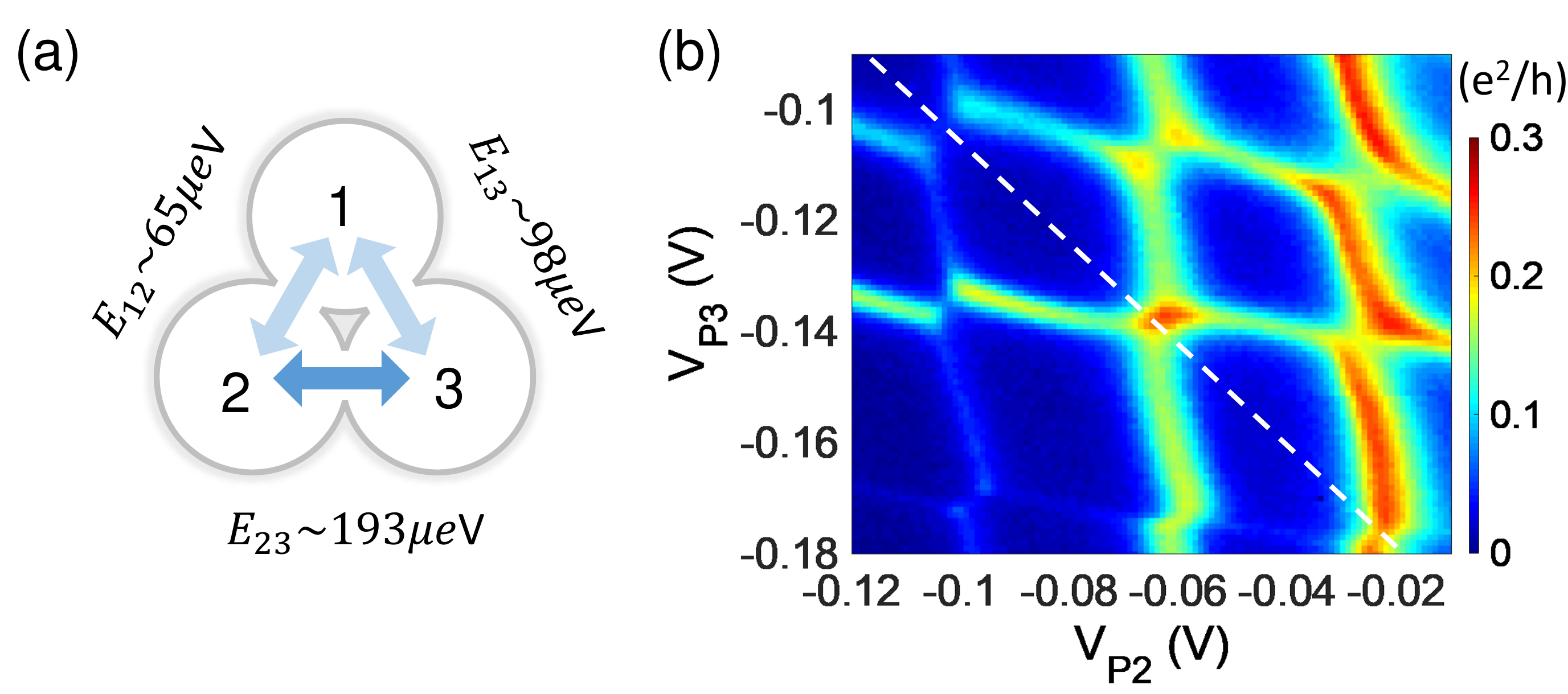}
		\caption{(color online) (a) Couplings between the QDs are set to opposite to the EP condition. (b) Measured conductance (G2+G3) through QD2 and QD3 in the case of (a). The white dashed line represents the charge degeneracy line of QD1 (separately measured). This line crosses the DQD degeneracy line of QD2 and QD3 at the center of the figure.}
		\label{fig:short}
		\end{center}
	\end{figure}

	A DQD degeneracy line is achieved, by tuning electrostatic coupling between the QDs to be $E_{12}+E_{13}<E_{23}$ as shown in Fig. 5(a). The coupling condition is opposite to that for the EP degeneracy shown in the Fig. 3(a), as the coupling between QD2 and QD3 dominates over the others. Under the condition, it is possible that a DQD degeneracy line of QD2 and QD3 crosses a charge degeneracy line of QD1 at almost right angles, showing a four-fold degeneracy point [see the center of the stability diagram measured by G2+G3 in Fig. 5(b)]. This is distinct from the EP degeneracy line that is parallel to a charge degeneracy line of QD1 [see Fig. 1(d) and Fig. 3]. To avoid Coulomb repulsion between QD2 and QD3, the effective charges of QD2 and QD3 must have the opposite sign to each other along the DQD degeneracy line, hence, the electron pairing cannot occur.
	
	In summary, we have shown that electron pairing, previously reported in a quadruple quantum dot \cite{Hamo2016}, can occur in a simpler system of a TQD. The electron pairing in two dots of the TQD is mediated by the third dot, when the third dot strongly couples with the other two via Coulomb repulsion. The transport spectroscopy, monitoring electron conductance through an individual QD, allows us to find the degeneracy for the pairing, the condition for the degeneracy, and nontrivial transport behavior by the paring.
	
	We expect that the understanding of the electron pairing will offer basic principles for designing a system useful for studying nontrivial correlation effects such as negative-$U$ Anderson impurities \cite{Hamo2016,Yoo2014,Taraphder1991,Koch2007}.

	We thank M. Heiblum for discussion and experimental support. We also thank Wonjin Jang and Myung Won Lee for technical assistants. This work was supported by the National Research Foundation (NRF) of Korea Grant funded by the Korean Government (MSIP) (NRF-2014R1A2A1A11053072), (NRF-2016R1A5A1008184; H.-S.S. and Y.C.), and KIST Institutional Program (Project No. 2E26681). The cryogenic measurement was done on the low temperature facility supported by Samsung Science and Technology Foundation under Project Number SSTF-BA1502-03. Y.C. was partly supported by the The Weston Visiting Professorships from Weizmann Institute of Science(IL).
\bibliographystyle{apsrev4-1}

\begin{thebibliography}{100}
\bibitem{Fit}
	The stability diagram is computed, using capacitances $C_{AB}$=33aF, $C_{BC}$=35aF, $C_{AC}$=10aF, $C_{A\alpha}$=5.2aF, $C_{B\beta}$=3.4aF, $C_{C\gamma}$=4.0aF, $C_{A\gamma}=C_{B\gamma}=C_{A\beta}=C_{C\beta}=C_{B\alpha}=C_{C\alpha}$=0.4aF, $C_{AR}$=176aF, $C_{BR}$=167aF, $C_{CR}$=187aF. We follow the notations of the capacitances in Ref. [6], and the subscript R stands for the 2DEG reservoir.
\end{thebibliography}
\end{document}